\documentclass[letterpaper]{JHEP3}

\usepackage{amssymb}
\usepackage{amsmath}
\usepackage{graphicx}
\usepackage{bm}

\newcommand{\bea}{\begin{eqnarray}}
\newcommand{\eea}{\end{eqnarray}}
\newcommand{\beq}{\begin{equation}}
\newcommand{\eeq}{\end{equation}}
\newcommand{\nn}{\nonumber}

\newcommand{\ignore}[1]{}

\def\<{\langle}
\def\>{\rangle}

\def\Hinf{H_{\rm inf}}
\def\cH{{\cal H}}
\def\HL{H_\Lambda}

\def\bfx{{\bm x}}

\def\bfk{{\bm k}}
\def\bfq{{\bm q}}
\def\bfp{{\bm p}}
\def\rCMB{r_{\!\rm cmb}}
\def\rCAH{r_{\!\rm eq}}
\def\rCP{r_{\!\rm cp}}

\def\A{{\cal A}}
\def\tr{\tilde{r}}

\def\dL{\delta^{\mbox{\tiny L}}}

\allowdisplaybreaks[4]

\title{The CMB and the measure of the multiverse}

\author{Michael P.~Salem\\
Stanford Institute for Theoretical Physics and Department of Physics, Stanford University, Stanford, California 94305, USA}

\abstract{In the context of eternal inflation, cosmological predictions depend on the choice of measure to regulate the diverging spacetime volume.  The spectrum of inflationary perturbations is no exception, as we demonstrate by comparing the predictions of the fat geodesic and causal patch measures.  To highlight the effect of the measure---as opposed to any effects related to a possible landscape of vacua---we take the cosmological model, including the model of inflation, to be fixed.  We also condition on the average CMB temperature accompanying the measurement.  Both measures predict a 1-point expectation value for the gauge-invariant Newtonian potential, which takes the form of a (scale-dependent) monopole, in addition to a related contribution to the 3-point correlation function, with the detailed form of these quantities differing between the measures.  However, for both measures both effects are well within cosmic variance.  Our results make clear the theoretical relevance of the measure, and at the same time validate the standard inflationary predictions in the context of eternal inflation.} 

\preprint{SU-ITP-12/13}

\begin{document}


\section{Introduction}
\label{sec:introduction}

Spacetime might feature eternal inflation \cite{Vilenkin:1983xq,Linde:1986fc}.  Theoretically, eternal inflation occurs about any metastable state with a decay rate smaller than its Hubble rate \cite{Guth:1982pn}, and such states are expected to exist in string theory \cite{Bousso:2000xa,Kachru:2003aw,Susskind:2003kw}.  Moreover, the spectrum of temperature fluctuations in the cosmic microwave background (CMB) is broadly consistent with expectations from an early period of slow-roll inflation \cite{Komatsu:2010fb}, and the simplest models of slow-roll inflation also exhibit eternal inflation (see for example \cite{Guth:2007ng}).  Observationally, supernova and CMB data indicate that the energy budget of our universe is dominated by some `dark energy' with equation of state $w\simeq -1$ \cite{Riess:1998cb,Perlmutter:1998np,Komatsu:2010fb}, and the simplest explanation of this involves a cosmological constant (vacuum energy density), which would also imply eternal inflation.    

Eternal inflation generates diverging spacetime volume.  Consequently, the pre-conditions to any type of measurement are established a diverging number of times, and to predict the relative probability of different possible outcomes of a measurement requires a procedure to regulate these divergences.  Such a regulation procedure is called a measure---the measure in effect constructs the probability space for physical observables in an infinite spacetime---and the `measure problem' refers to the fact that different seemingly sensible measure proposals sometimes give dramatically different physical predictions (for recent reviews of the measure problem, see for example \cite{Freivogel:2011eg,Salem:2011qz}).
     
The choice of measure in principle affects all probabilistic predictions.  This includes, in particular, any predictions related to the spectrum of inflationary perturbations.  Of course, insofar as the standard formulation of inflationary predictions (see for example \cite{Mukhanov:1990me}) has met with empirical success, the choice of measure should make predictions at least approximately in agreement.  Nevertheless, the standard formulation cannot stand by itself, as an arbitrarily precise and self-contained framework, as it relies on assumptions that require justification in the context of eternal inflation.  It is worthwhile to elaborate on this point.  

Given suitable initial conditions, inflation generates a spacetime region with approximate Friedmann-Robertson-Walker (FRW) symmetries.  These symmetries are broken by quantum fluctuations, which decohere to represent an ensemble of classical perturbations, with branching ratios given by projection onto the Bunch-Davies vacuum \cite{Bunch:1978yq}.  The branching ratios are related to probabilities via the usual Born rule, but these are probabilities for perturbations, not the outcomes of measurements.  The standard formulation makes predictions by assuming that measurements are equally likely to be performed at any comoving coordinate in the FRW geometry.  (Put another way, it assumes that there are no correlations between the locations of measurements and the perturbations that they measure.)  Actually, this assumption is not exactly valid even in a finite spacetime with global FRW symmetries.  For example, a subset of the ensemble of inflationary perturbations is, by chance, homogeneous at a level far below the variance characterizing the entire ensemble.  Elements of this subset feature stunted structure formation and therefore relatively few observers to measure them.  Therefore, perturbations in this subset should contribute less to expectation values than the above assumption would indicate.  This `anthropic' effect is very small, and ignoring it can be viewed as a convenient approximation.  Anthropic effects notwithstanding, the above assumption would be well motivated if the global spacetime were finite (and nearly all measurements occurred in the FRW region) or if the FRW symmetries described the global spacetime.  However, neither of these conditions holds in the context of eternal inflation.

To illustrate how the choice of measure could invalidate the above assumption, consider as an analogy the selection of random points via throwing darts at a wall.  If the wall has a hemispherical bulge, an observer on the bulge who is small compared to the bulge might recognize the local O(3) symmetry and suppose that her random position is selected according to that.  This observer would be incorrect.  This situation is not unlike that of the worldline-based measures studied in this paper, with the darts representing the future histories of the worldline, the wall representing the reference frame of the worldline, and the bulge representing a frame with local FRW symmetries.  Here the effect of the `measure' is large in the sense that random points are much less likely to lie where the tangent to the bulge is parallel to the trajectories of the darts.  On the other hand, insofar as the local O(3) symmetry implies that a local environment is independent of its coordinates on the bulge, the observer cannot discern her (likely) special location.  Nevertheless, the same effect applies with respect to any small perturbation on the bulge, though now the size of the effect is suppressed in terms of the smallness of the perturbation.  This foreshadows the conclusions of our analysis.  

We study the spectrum of inflationary perturbations in the context of two measures, the fat geodesic measure \cite{Bousso:2008hz} (see also \cite{Nomura:2011dt,Larsen:2011mi}) and the causal patch measure \cite{Bousso:2006ev}.  We focus on these measures in part because we find their formulations to be particularly amenable to the analysis of inflationary perturbations, and in part because these measures are among a small set known to pass three important phenomenological tests.  In particular, they do not possess a youngness problem \cite{Tegmark:2004qd,Guth:2007ng,Bousso:2007nd},\footnote{These measures still feature what might be called a youngness pressure:  all else being equal, they assign greater weight to sequences of events that span less time.  However, this effect is comparatively small---roughly speaking, it is only significant for sequences of events that span a Hubble time, and therefore it does not pose any obvious problems for phenomenology---though its theoretical implications are a matter of debate \cite{Bousso:2010yn,Guth:2011ie}.} nor do they feature the $Q$ or $G$ catastrophes \cite{Feldstein:2005bm,Garriga:2005ee,Graesser:2006ft}, and they can avoid Boltzmann-brain domination \cite{Dyson:2002pf,Albrecht:2004ke,Page:2006dt,Bousso:2006xc,Bousso:2011aa}, given plausible assumptions about decay rates in the landscape \cite{Bousso:2006xc,Bousso:2008hz}.  (It should be noted that the causal patch measure assigns an overwhelming probability to observe a negative vacuum energy \cite{Salem:2009eh,Bousso:2010zi}.  This is a serious issue, but one might speculate that it can be resolved without affecting predictions for positive-energy vacua such as ours.  The causal patch measure might also predict too small of a positive vacuum energy, depending on the details of the landscape \cite{Bousso:2010zi}.)  The causal patch measure is closely related to the lightcone-time cutoff measure \cite{Bousso:2009dm,Bousso:2010id}, and the assumptions of our analysis establish a condition by which these measures make identical predictions \cite{Bousso:2009mw}.  The fat geodesic measure is closely related to the scale-factor cutoff measure \cite{DeSimone:2008bq,DeSimone:2008if}, but the correspondence is not precise \cite{Bousso:2008hz}, and we expect these measures to give different predictions at our level of analysis.  The CAH+ measure \cite{Vilenkin:2011yx,Salem:2011mj} has some similarities with the lightcone-time cutoff measure, but again the relationship is not precise, and we expect these measures to make different predictions at our level of analysis.  The `equilibrium' measure \cite{Vanchurin:2012xm} has some similarities with the fat geodesic measure, however the two also have important differences and so we cannot assert that they make the same predictions here.

To highlight the effects of the measure---as opposed to any effects related to performing cosmological measurements in the context of a landscape of vacua---we take the cosmological model, including the model of inflation, to be specified.  We also condition on a given average CMB temperature at the point of measurement.  (To keep the analysis simple, we only set this condition in an approximate manner, however we do not expect this to affect the qualitative form of the results.)  We find that both measures predict a `1-point' expectation value for the Fourier components of the gauge-invariant Newtonian potential $\Psi$, this taking the form of a scale-dependent monopole.  Accordingly, they both predict a contribution to the expectation value of the 3-point correlation function, when one of the three insertions of $\Psi$ is a monopole.  The predictions of both measures have the same order of magnitude, but the precise size and form of the predictions differ between the measures.  With the fat geodesic measure, the effect comes in part from the tendency of the worldline that defines the measure to gravitate toward over-densities, these over-densities correlating with $\Psi$.  With the causal patch measure, there is a similar effect, and in addition there is an effect coming from the correlation between the size of the causal patch---which affects the number of measurements that are performed in a given causal patch---and $\Psi$.  While the monopole is just a constant background with respect to the primary CMB perturbations, it appears in secondary effects, including the angular size of acoustic peaks (see for example \cite{Hu:1997mn}).  Nevertheless, we find the sizes these measure effects to be well within cosmic variance.    

While our work demonstrates the measure dependence of the inflationary spectrum, the smallness of the effects can be seen as validation of the standard formulation of inflationary predictions.  Yet, it is important to keep in mind that while the fat geodesic and causal patch measures have achieved some phenomenological successes, no measure has yet to acquire a broadly compelling theoretical motivation.  Hence, the validation of the standard formulation should be seen more as a proof of principle.  Nevertheless, familiarity with the phenomenology of measures lends one to suspect that any measure that survives the three major phenomenological tests listed above (for instance the scale-factor cutoff measure and CAH+ measure) will permit similar conclusions with respect to inflationary observables.

Before proceeding, we note that an effect on the inflationary spectrum coming from the proper-time cutoff measure \cite{Linde:1993nz,Vilenkin:1994ua} was previously discussed in \cite{Linde:1994gy}.  Compared to our analysis, that work is less explicit in the size of the effect (which also includes a scale-dependent 1-point expectation value in the form of a monopole).  Also, the proper-time cutoff measure is known to exhibit the youngness problem as well as the $Q$ and $G$ catastrophes mentioned above, and is therefore no longer considered as a viable measure on eternal inflation.    

The remainder of this paper is organized as follows.  In Section \ref{sec:measures}, we briefly review the fat geodesic and causal patch measures, and then set up the general framework for predicting inflationary observables in each case.  For concreteness we introduce a number of conditional assumptions; these are sufficient but not necessary to demonstrate the measure dependence.  Section \ref{sec:model} describes our model assumptions, which are intended to provide a simple approximation of the standard cosmological model.  In Section \ref{sec:FG}, we compute the expectation values of correlation functions of $\Psi$ in the context of the fat geodesic measure, while in Section \ref{sec:CP} we perform the same calculations but for the causal patch measure.  Finally, we summarize our results and draw our conclusions in Section \ref{sec:conclusions}.


\section{Fat geodesic and causal patch measures}
\label{sec:measures}

\subsection{Brief review}

We first review the fat geodesic and causal patch measures.  Both of these measures focus on the future histories surrounding a worldline emanating from some initial conditions, taking for granted a probability space defined over the set of possible initial conditions.  It is assumed that for practical purposes, these future histories can be represented by an ensemble of semi-classical spacetimes.  Different elements of this ensemble correspond to different Coleman--De Luccia (CDL) bubbles \cite{Coleman:1980aw,Brown:2007sd} nucleating in different places, different configurations of decohered inflaton fluctuations, etc.  Eventually, the worldline encounters a CDL bubble with negative vacuum-energy density and is inevitably drawn into a big crunch singularity, at which point the worldline is terminated.  (Some worldlines never encounter such a singularity, but these form a set of measure zero in the fat geodesic and causal patch measures.)  The fat geodesic measure only counts events that occur within some small, fixed physical distance orthogonal to the worldline (the `fat geodesic'), whereas the causal patch measure only counts events that occur within the past lightcone of the end of the worldline (the `causal patch').  We denote the ensemble of causal patches $\Sigma$, and take this to define an ensemble of fat geodesics as well, in the latter case simply restricting attention to the spacetime within the fat geodesic.  Each element $i$ of $\Sigma$ is assigned a weight ${\cal I}_i$ corresponding to the quantum-mechanical branching ratio to the semi-classical spacetime that it represents.  

The fat geodesic and causal patch measures use their corresponding rules to assign relative probabilities to classes of local events in the semi-classical spacetime.  Specifically, the relative probability assigned to some event of type $E$ in the fat geodesic measure is
\beq
P(E) \propto \sum_{i\in\Sigma}\,\, {\cal I}_i\,{\cal N}_i(E) \,,
\eeq
where ${\cal N}_i(E)$ is the number of events of type $E$ within the fat geodesic of $i$.  In the causal patch measure, this relative probability is
\beq
P(E) \propto \sum_{i\in\Sigma}\,\,{\cal I}_i\,\overline{\cal N}_{\!i}(E) \,,
\eeq
where $\overline{\cal N}_{\!i}(E)$ is the number of events $E$ within the causal patch $i$.  One is usually interested in predicting conditional probabilities, in which case one only counts events when they satisfy the prescribed conditions, and one can likewise restrictthe ensemble of causal patches $\Sigma$ to the subset containing events satisfying those conditions.

\subsection{Assumptions and approximations}
\label{ssec:aa}

These probabilities can be made more concrete with some plausible assumptions about the landscape.  First of all, when making predictions for observers like us, we can condition on our extremely small (in Planck units) and therefore extremely rare (in the landscape) vacuum-energy density.  Thus it can be assumed that no vacua with smaller-magnitude vacuum-energy densities are nearby in the landscape configuration space.  Since transitions to higher-energy vacua are exponentially suppressed next to transitions to lower-energy vacua, this means that among the set of worldlines that enter our vacuum, the overwhelming majority transition next to a negative-energy vacuum, after which the worldline quickly terminates.  Therefore, to a good approximation we can restrict attention to the sub-ensemble of $\Sigma$ in which the worldline encounters our vacuum only once, terminating soon after our vacuum decays.  Since the decay of our vacuum is itself likely to be strongly exponentially suppressed (in Hubble units), the causal patch of the worldline can be approximated as the past lightcone of future infinity, for this purpose treating our vacuum as if it does not decay.  
  
Moreover, to demonstrate the measure dependence of inflationary observables, it is sufficient to condition on a specific model of the local cosmology.  In particular, we assume that the observable universe is contained in a CDL bubble, which nucleated many Hubble times after the nucleation of its progenitor bubble, and for which the effects of bubble-wall perturbations and collisions with other bubbles are negligible.  We assume there is a finite period of slow-roll inflation within the bubble to erase its initial spatial curvature.  (This period of inflation generates the inflationary observables.)  The particular model of inflation, as well as the model describing the subsequent cosmological evolution, is taken to be completely fixed.  We even condition on a given value of the average CMB temperature, which we denote $T_{\rm obs}$, at the point of measurement.\footnote{We could forego this condition, but this would invite us to condition more explicitly on anthropic criteria, which at this stage we prefer to avoid (we discuss this and related issues in Section \ref{sssec:conditions}).}   

\begin{figure*}[t!]
\begin{center}
\begin{tabular}{ccc}
\includegraphics[width=0.33\textwidth]{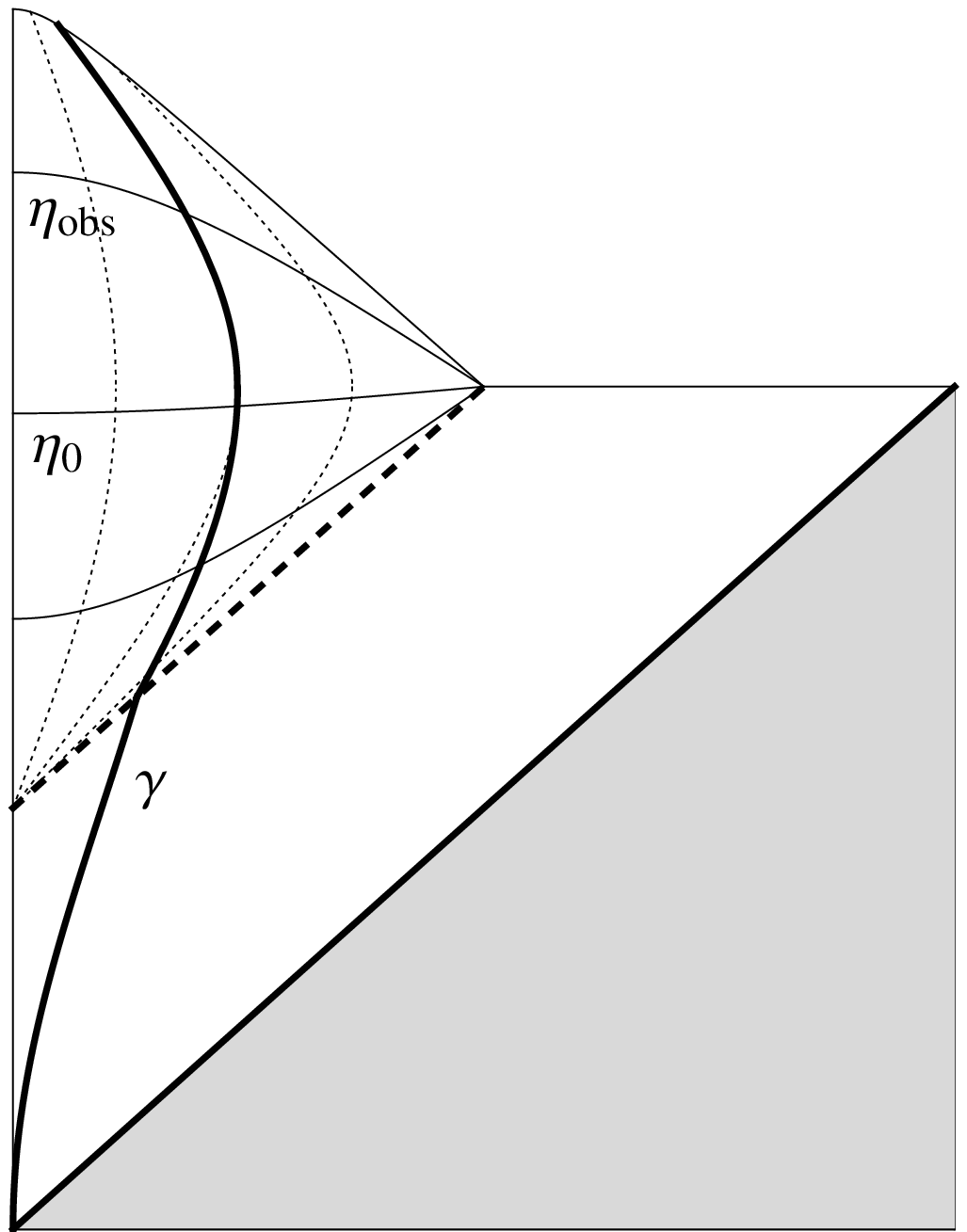} 
& \phantom{spacespa} &
\includegraphics[width=0.33\textwidth]{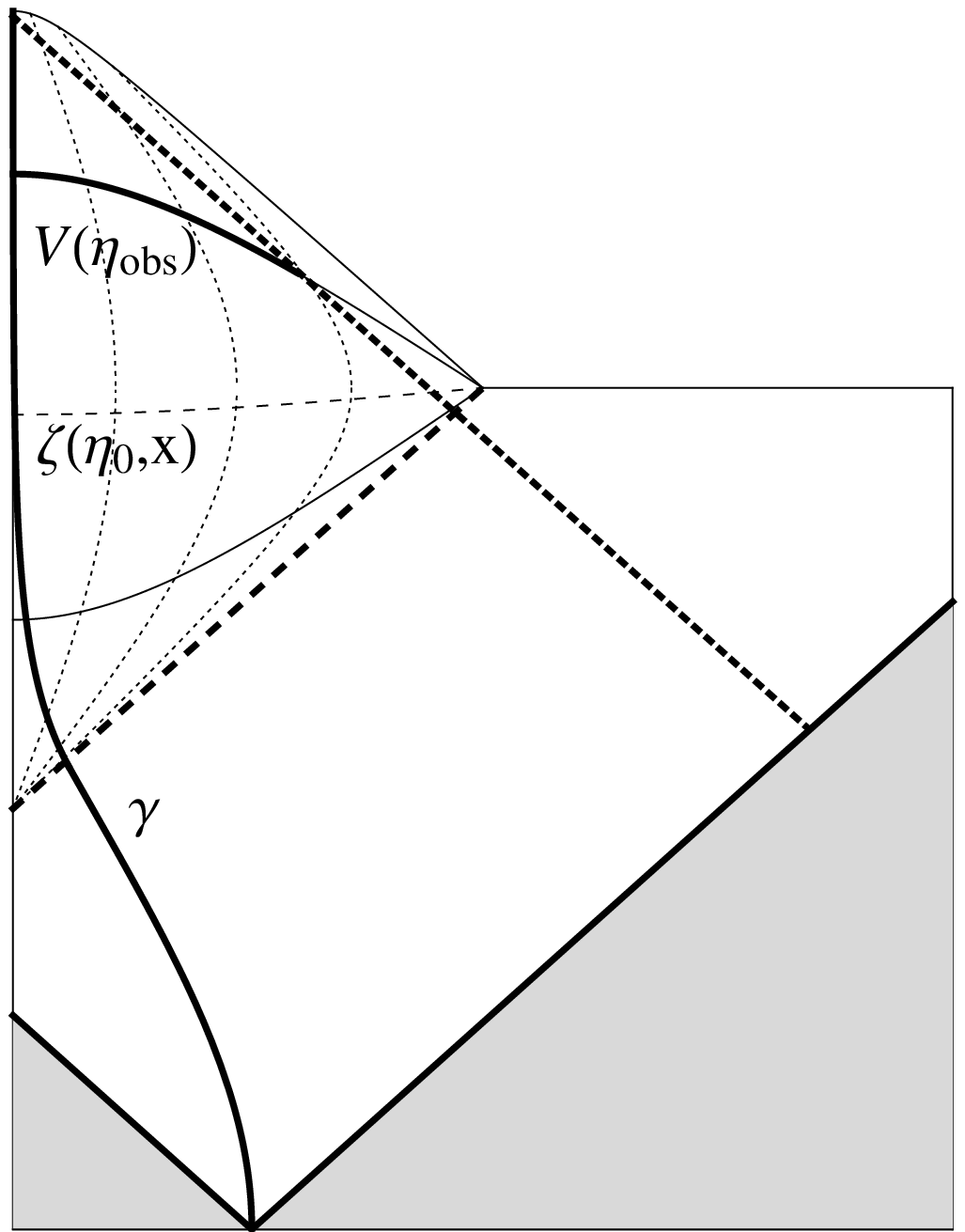} 
\end{tabular}
\caption{\label{fig:bubble}  Left:  toy conformal diagram of a comoving (with respect to the spatially flat de Sitter chart) worldline $\gamma$ (thick curve) entering a CDL bubble (thick dashed line), in which it is boosted with respect to the comoving open-FRW chart (solid and dotted curves), but quickly becomes comoving during inflation.  The diagram indicates the times $\eta_0$ and $\eta_{\rm obs}$ referred to in the text.  One should imagine a small negative-vacuum-energy CDL bubble at the tip of the worldline.  Right:  FRW symmetry in the bubble has been exploited to make $\gamma$ asymptote toward the center of the bubble.  The causal patch of the worldline is indicated (thick dotted curve), as are the three-volume of the causal patch at $\eta_{\rm obs}$ (thick curve) and the curvature perturbation at $\eta_0$ (dashed curve).}
\end{center}
\end{figure*}

Figure \ref{fig:bubble} provides two equivalent conformal diagrams of the geometry.  Our assumptions allow us to focus on a single bubble embedded in the spatially flat de Sitter chart.  The interior of the bubble is an infinite, open-FRW universe.  The worldline defining either measure can enter the bubble at any point along the bubble wall, and for entry points displaced from the center of the bubble the worldline will be boosted with respect to the open-FRW chart.  Nevertheless, this boost is redshifted as the scale factor within the bubble expands, and the worldline quickly becomes comoving in the open-FRW frame.  To simplify visualization, one can then exploit the FRW symmetries in the bubble to boost the worldline so that it appears to have come from the origin of coordinates in the bubble.

\subsection{Formal application to inflationary observables}

Consider the decohered curvature perturbation $\zeta$ on some fixed-FRW-time hypersurface $\eta=\eta_0$ near the end of inflation in some bubble like ours (a gauge-invariant specification of $\zeta$ is given in Section \ref{ssec:gauge}).  Given our assumptions, the (relevant) future evolution of the bubble depends only on the specific configuration of this (classical) perturbation.  Therefore, we can label fat geodesics / causal patches according to their particular perturbation $\zeta(\eta_0,\bfx)$, taking the worldline defining each measure to pass through $\bfx=0$.  The branching ratio ${\cal I}_{\zeta}$ is then just the quantum-mechanical branching ratio to create the specified perturbation $\zeta$.  To make this precise, note that our assumptions allow us to work in the effective Fock space $\{|\zeta\>\}$ of curvature perturbations on an open--de Sitter chart (the usual `Fock space' of cosmological perturbation theory in the single-bubble background \cite{Bucher:1994gb,Sasaki:1994yt,Garriga:1998he}).  Then we can write the branching ratio
\beq
{\cal I}_{\zeta} = |\<\zeta|0\> |^2 \,,   
\eeq
where $|0\>$ denotes the Bunch-Davies vacuum.\footnote{Distinct configurations of $\zeta$ will sometimes correspond to the same fat geodesic / causal patch, in part because the positive cosmological constant in our bubble implies that the defining worldline is in causal contact with only a finite portion of the infinite $\eta=\eta_0$ hypersurface.  This degeneracy would be relevant to computing the branching ratios to distinct fat geodesics / causal patches if the field configuration within the causal patch were not statistically independent of the field configuration outside of the causal patch.  Our adoption of the Bunch-Davies vacuum guarantees the requisite statistical independence, however this is ultimately an assumption about the probability space over initial conditions in these measures.}  

To better understand the counting of events with ${\cal N}_{\zeta}$ and $\overline{\cal N}_{\!\zeta}$, it will help to refine our notation.  The spectrum of inflationary perturbations seen by a given observer contains a lot of information, and one may wish to characterize it with some statistic $z$.  This statistic is a functional of the particular curvature perturbation $\zeta$ within the past lightcone of the observer, who resides at some point $\{\eta,\bfx\}$.  Thus we should write $z=z[\eta,\bfx,\zeta]$.  (We emphasize that $z$ is a statistic characterizing the data set of a single observer at a specific location in a particular semi-classical spacetime.)  Let $E_{z\to\tilde{z}}$ denote the event of measuring $z$ to attain the value $\tilde{z}$, regardless of where the measurement is performed or what is the specific perturbation $\zeta$.  The number of such events in a given fat geodesic is
\beq
{\cal N}_{\zeta} (E_{z\to\tilde{z}}) = 
\int d\eta \int_{{\rm FG}[\zeta]}\! \sqrt{-g[\eta,\bfx,\zeta]}\,d\bfx\,
\delta\big[\eta-\eta_{\rm obs}(\bfx)\big]\, 
\delta\big\{z\big[\eta,\bfx,\zeta\big]-\tilde{z}\big\}\,
\A\big[\eta,\bfx,\zeta\big] \,,
\label{FGdensity}
\eeq
where the inner integral is understood to cover the fixed physical three-volume within the fat geodesic at time $\eta$, the first delta function selects for when the average CMB temperature is $T_{\rm obs}$---with $\eta_{\rm obs}(\bfx)$ being the time at which this happens according to a comoving geodesic through $\bfx$---the second delta function selects for when the statistic $z$ gives $\tilde{z}$, and $\A[\eta,\bfx,\zeta]$ gives the probability for there to be a measurement at $\{\eta,\bfx\}$, which depends on $\zeta$.  

We have set up the calculation so that the worldline defining the fat geodesic measure is comoving and passes through the coordinate $\bfx=0$ at some time $\eta_0$ near the end of inflation.  Nevertheless, this worldline will subsequently be drawn toward over-densities.  Therefore, the inner integral in (\ref{FGdensity}) is not centered at $\bfx=0$, but instead follows some trajectory that is correlated with the particular perturbation $\zeta$ through which the worldline evolves.  Insofar as an initially uniform distribution of comoving worldlines is eventually distributed according to the cold dark matter density, we can account for this effect by including a factor of the matter density $\rho_{\rm m}[\eta,\bfx,\zeta]$ in $\A$, after which we can again take the defining worldline to be centered at $\bfx=0$.  Note that $\rho_{\rm m}$ is a gauge-dependent quantity.  Since we are conditioning on a given average CMB temperature $T_{\rm obs}$, we take $\rho_{\rm m}$ to be specified with respect to a foliation on which hypersurfaces of fixed $\eta$ see fixed average CMB temperature.  Meanwhile, we take the probability for there to be a measurement to be proportional to the matter density, so that $\A$ contains an additional factor of $\rho_{\rm m}$.  For simplicity we do not distinguish between cold dark matter and baryonic matter.  Meanwhile, the physical three-volume within the fat geodesic is intended to be very small---in particular much smaller than the scales of non-linear structure formation---which means the integrals and delta functions of (\ref{FGdensity}) essentially select for fat geodesics for which $z[\eta_{\rm obs}(0),0,\zeta]=\tilde{z}$.

Putting all of this together, we write
\beq
P(E_{z\to\tilde{z}}) \propto \int [d\zeta]\,|\<\zeta|0\>|^2
\,\rho_{\rm m}^2\big[\eta_{\rm obs}(0),0,\zeta\big]\,
\delta\big\{z\big[\eta_{\rm obs}(0),0,\zeta\big]-\tilde{z}\big\} \,.
\label{FGP2}
\eeq
We interpret $P(E_{z\to\tilde{z}})$ as the probability distribution for measured values of the observable $z$.  Accordingly, we write the (semi-classical) expectation value
\beq
\<z\>_{\rm FG} = \frac{1}{N}\int d\tilde{z}\,\tilde{z}\,P(E_{z\to\tilde{z}}) \,, 
\label{FGexp}
\eeq
where
\beq 
N\equiv \int d\tilde{z}\,P(E_{z\to\tilde{z}}) \,.
\eeq
While the above analysis is semi-classical, it is illuminating to express the result in a more quantum-mechanical language.  Consider the promotion of the classical observables $z$ and $\rho_{\rm m}$ to quantum operators $\hat{z}$ and $\hat{\rho}_{\rm m}$ (on the effective Fock space $\{|\zeta\>\}$), according to replacing their dependence on $\zeta$ with a dependence on $\hat{\zeta}\equiv |\zeta\>\<\zeta|\,\zeta$.  Then we can write
\bea
\<z\>^{\phantom{'}}_{\rm FG} &=& \frac{1}{N} \int d\tilde{z} 
\int [d\zeta]\,\delta\big\{z\big[\eta_{\rm obs}(0),0,\zeta\big]
-\tilde{z}\big\}\,\<\zeta|0\>\<0|\hat{\rho}_{\rm m}^2\hat{z}|\zeta\> 
\quad\,\,\\
&=& \frac{1}{N}\,\<0|\hat{\rho}_{\rm m}^2\hat{z}|0\> \,, 
\label{FGresult}
\eea
where now $N=\<0|\hat{\rho}_{\rm m}^2|0\>$.

We emphasize that we have not actually solved a quantum-mechanical problem; we have merely found an effective Fock space $\{|\zeta\>\}$ on which certain operators reproduce the predictions of the fat geodesic measure in a semi-classical picture of the multiverse.  The prediction is apparently different than that of the standard approach.  In the standard formulation, the expectation value for $z$ is $\<z\>^{\phantom{'}}_{\rm SM}=\<0|\hat{z}|0\>$.  The result (\ref{FGresult}) differs because $\rho_{\rm m}$ depends perturbatively on $\zeta$, whose statistics are being assessed by $z$.  Essentially, the fat geodesic measure enhances the probability to observe curvature perturbations $\zeta$ that increase the matter density at the origin (the location of the defining worldline), as this increases the likelihood that the fat geodesic includes a measurement of $\zeta$.  There is both an anthropic effect, as we assume measurements are performed in proportion to the matter density, and what might be called a measure selection effect, due to the tendency of over-densities to attract the fat geodesic.  Of course, the correlation between $\rho_{\rm m}[\eta_{\rm obs}(0),0,\zeta]$ and $z[\eta_{\rm obs}(0),0,\zeta]$ is very small.  Our goal is merely to identify the main effect of the measure.
     
The density of events $E_{z\to\tilde{z}}$ in the causal patch measure is 
\beq
\overline{\cal N}_{\!\zeta}(E_{z\to\tilde{z}}) 
\propto \int d\eta\int_{{\rm CP}[\zeta]}\! 
\sqrt{-g[\eta,\bfx,\zeta]}\,d\bfx\,
\delta\big[\eta-\eta_{\rm obs}(\bfx)\big]\, 
\delta\big\{z\big[\eta,\bfx,\zeta\big]-\tilde{z}\big\}\,
\A\big[\eta,\bfx,\zeta\big] \,,
\label{CPdensity}
\eeq
where the inner integral is understood to run over the three-volume in the causal patch at time $\eta$ (which we emphasize is a functional of the curvature perturbation $\zeta$), and the terms in the integrand play the same roles as before.  The quantity $\overline{\cal N}_{\!\zeta}$ is superficially very similar to ${\cal N}_\zeta$, but the integration over the causal patch---which is comparable to the Hubble volume in size---as opposed to an integration over the fat geodesic---which is much smaller than the scales of non-linear structures---creates a crucial difference.  The argument concerning $\A$ for the fat geodesic measure applies here, except while the defining worldline is boosted to $\bfx=0$, any given observer is not.  Therefore
\beq
\A[\eta,\bfx,\zeta] \propto \rho_{\rm m}\big[\eta,0,\zeta\big]\,\rho_{\rm m}\big[\eta,\bfx,\zeta\big] \,.
\eeq   
As before, it is illuminating to promote the various quantities to quantum operators, but here we should be mindful of the dependence on $\bfx$.  For the moment we simply write 
\beq
\hat{Z} \equiv \int d\eta\int_{{\rm CP}[\zeta]}\! 
\sqrt{-g[\eta,\bfx,\zeta]}\, d\bfx\,
\delta\big[\eta-\eta_{\rm obs}(\bfx)\big]\,
\A\big[\eta,\bfx,\zeta\big]\,
z\big[\eta_{\rm obs}(\bfx),\bfx,\zeta\big]\Big|_{\zeta\to\hat{\zeta}} 
\,\,. \label{Zdef}
\eeq
Then, following the steps outlined above for the fat geodesic measure, we obtain 
\beq
\<z\>^{\phantom{'}}_{\rm CP} = \frac{1}{\overline{N}} 
\int [d\zeta]\, \<\zeta|0\>\<0|\hat{Z}|\zeta\> 
= \frac{1}{\overline{N}}\,\<0|\hat{Z}|0\> \,,
\label{CPprediction} 
\eeq
where the normalization factor is now $\overline{N}=\<0|\hat{Z}|_{z\to 1}|0\>$.

We reiterate that we have not actually solved a quantum-mechanical problem; we have merely identified certain operators that, when applied to the effective Fock space, reproduce the predictions of the causal patch measure in a semi-classical picture of the multiverse.  Note that the expectation value (\ref{CPprediction}) for the causal patch measure is different than the expectation value (\ref{FGresult}) for the fat geodesic measure.  In particular, the integrations over the causal patches in $\hat{Z}$ and $\hat{Z}|_{z\to 1}$ do not factor out and cancel in the ratio of (\ref{CPprediction}), in part because the volume of the causal patch depends on $\zeta$ and in part because $\A$ and $z$ depend on the position in the causal patch.  The fat geodesic measure assigns a higher weight to curvature perturbations $\zeta$ that increase the matter density at the origin, as this increases the likelihood for the fat geodesic to enclose a measurement of $\zeta$, while the causal patch measure in addition assigns a higher weight to curvature perturbations that generate a larger causal patch, as this encloses more measurements elsewhere in the causal patch.  Granted, in each case the measure dependence on $\zeta$ is perturbative in $\zeta$, so to a good approximation the two measures make the same inflationary predictions.  Our goal is merely to identify the leading order measure dependence.


\section{Toy cosmological model}
\label{sec:model}

\subsection{FRW cosmology}
\label{ssec:model}

We continue to adopt the assumptions and approximations of Section \ref{ssec:aa}.  Correspondingly, we take the observable universe to be contained in a CDL bubble.  On small comoving scales the line element has negligible spatial curvature, and at leading order can be written
\beq
ds^2 = a^2(\eta)\left( -d\eta^2 + d\bfx^2 \right) ,
\label{FRWmetric}
\eeq
where $d\bfx^2$ is the three-dimensional Cartesian line element.  In this section we describe the FRW cosmology.  Most of the details are unimportant, and our primary interest is to establish some notation and understand the relative sizes of important comoving scales.  

For the purpose of understanding the background geometry, we model inflation as essentially vacuum-energy domination, and denote the (constant) Hubble rate $\Hinf$.  After several $e$-folds of inflation, the scale factor can be written
\beq
a(\eta) = -\frac{1}{\Hinf\,(\eta+\eta_\infty)} \,,
\eeq
where $\eta_\infty$ is an integration constant, which we set so that $\eta$ asymptotically approaches zero (from below) during the late-time cosmological-constant dominated phase in our bubble.  We adopt this convention so that the comoving radius of the causal patch is given simply by $|\eta|$.  Note that during the early-time period of inflation, $\eta$ asymptotes toward $-\eta_\infty$.    

We assume that inflation gives way directly to a period of radiation domination, followed by matter domination, followed by cosmological-constant domination.  In each case we approximate the transition as instantaneous, solving the FRW field equations on each side of the transition by assuming the energy density behaves as a perfect fluid with the appropriate equation of state:  $w\equiv p/\rho =1/3$ during radiation domination, $w=0$ during matter domination, and $w=-1$ during cosmological-constant domination (here $p$ is the pressure and $\rho$ is the energy density).  The integration constants are set so as to make the scale factor and its first time derivative continuous across the transition.  The details are exactly the same as in \cite{Salem:2011mj}, and below we simply summarize the relevant results.

\begin{figure*}[t!]
\begin{center}
\begin{tabular}{cc}
\includegraphics[width=0.475\textwidth]{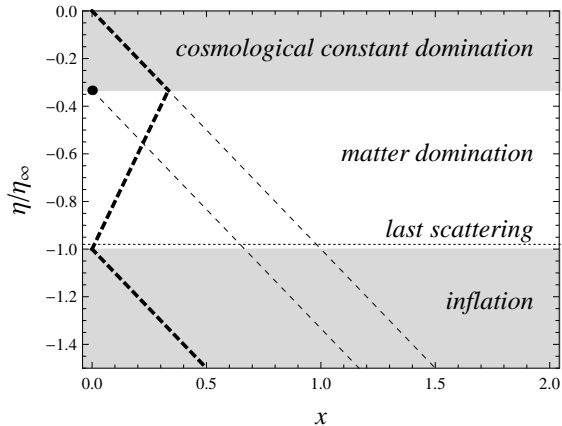} &
\phantom{sp}
\end{tabular}
\caption{\label{fig:history}  Zoom-in on a conformal diagram illustrating the local cosmology in our bubble (to scale).  The horizontal dotted line indicates the time of last scattering; the periods of slow-roll inflation and cosmological-constant domination are shaded gray.  The light dashed lines indicate the comoving sizes of the boundary of the causal patch and the past lightcone of an observer at time $\eta_{\rm obs}=-(1/3)\,\eta_\infty$.  The thick dashed line is the comoving size of the apparent horizon.}
\end{center}
\end{figure*}

Figure \ref{fig:history} illustrates the important cosmological transitions as a function of $\eta$.  At the end of inflation / onset of radiation domination, $\eta$ is exponentially close to $-\eta_\infty$.  The fractional change in $\eta$ during radiation domination is negligible compared to the change in $\eta$ during the subsequent matter domination, and by the end of matter domination we have $\eta=-(1/3)\,\eta_\infty$.  (In writing this and subsequent results, we have solved for the conformal-time evolution all the way to future infinity, so as to determine $\eta_\infty$ in terms of the other model parameters, and then have used that result to express the quantities of interest in terms of $\eta_\infty$, as opposed to in terms of the other model parameters.)  Surfaces of constant average CMB temperature correspond to surfaces of constant $\eta$, and to avoid a proliferation of scales we assume that measurements are performed at matter / cosmological-constant equality,
\beq
\eta_{\rm obs} = -(1/3)\,\eta_\infty \,.
\eeq
As for the scale-factor evolution of this toy model, it is only important to note the behavior after cosmological-constant domination,
\beq
a(\eta) = -\frac{1}{\HL\eta} \,, \qquad
\label{sf}
\eeq
where $\HL$ is the (constant) Hubble rate during cosmological-constant domination.

The `integration constant' $\eta_\infty$ sets the scale for all of the important comoving distances in this cosmology.  While our results do not depend on the actual size of $\eta_\infty$, we note that it can be written $\eta_\infty = 3\sqrt{\Omega_{\rm c}(\eta_{\rm obs})}$, where $\Omega_{\rm c}$ is the usual curvature parameter.  The comoving radius of the causal patch at the time of observation, $\rCP$, is simply given by the magnitude of the conformal time at the time of observation, that is
\beq
\rCP = |\eta_{\rm obs}|=(1/3)\,\eta_\infty \,.
\eeq
We also take interest in the comoving size of the surface of last scattering $\rCMB$, which is given by the comoving size of the observer's past lightcone at the time of last scattering.  Since the change in conformal time between the end of inflation and the time of last scattering is very small compared to $\eta_\infty$, this can be approximated 
\beq 
\rCMB = (2/3)\,\eta_\infty \,.
\eeq
Note that $\rCMB$ is of the same order as $\rCP$.  Finally, we also refer to the comoving size of the apparent horizon at matter / radiation equality, which we denote $\rCAH$.  For our purposes, it is only important that $\rCAH$ is very small compared to $\rCP$, but to be more precise we write
\beq
\rCAH = \big[(2/27)\,\HL\tau_{\rm eq}\big]^{1/3}\,\eta_\infty \,,
\eeq
where $\tau_{\rm eq}$ denotes the proper time (defined with respect to the time of reheating) of matter / radiation equality.  The relative size of $\rCAH$ and $\rCP$ is at the percent level.

\subsection{Cosmological perturbation theory}
\label{ssec:gauge}

We assume that metric perturbations are sourced entirely by a single inflaton field, we ignore tensor perturbations, and we focus on scales that are small compared to the curvature radius of our CDL bubble.  The most general line element can then be written \cite{Mukhanov:1990me}
\bea
ds^2 &=& a^2(\eta) \Big\{ -\!\big[1+2\phi(\eta,\bfx)\big] d\eta^2 
+2\,\partial_i B(\eta,\bfx)\,d\eta\,dx^i 
+ \big[1-2\psi(\eta,\bfx)\,\delta_{ij} \big. \Big.\nn\\ 
& & + \Big. \big. \,
2\,\partial_i\partial_j E(\eta,\bfx) \big] dx^i\,dx^j \Big\} \,.
\label{metric2}
\eea
Among $\phi$, $B$, $\psi$, $E$, and the inflaton perturbation $\delta\varphi$, there is actually only one scalar degree of freedom; the others are related to diffeomorphism (gauge) invariance.  In particular, under the generic `scalar' infinitesimal coordinate transformation
\beq 
\eta\to\eta+\sigma(\eta,\bfx) \,, \quad 
\bfx\to \bfx +\bm{\partial}\xi(\eta,\bfx) \,,
\label{redefinitions}
\eeq
the metric perturbations transform according to 
\beq
\phi\to\phi-\dot{\sigma}-\cH\sigma\,, \quad\,
B\to B+\sigma-\dot{\xi}\,, \quad\,
\psi\to\psi+\cH\sigma\,, \quad\,
E\to E-\xi \,,
\label{newgauges}
\eeq
where $\cH\equiv\dot{a}/a$, and dots denote derivatives with respect to $\eta$.  To help transform from one gauge to another, one can refer to the gauge-invariant variables  
\beq
\begin{array}{ll}
\Phi \equiv \phi + \cH\big(B-\dot{E}\big)+\dot{B}-\ddot{E} 
\phantom{\Big[\Big]} \quad\,\, &
\Psi \equiv \psi - \cH\big(B-\dot{E}\big) 
\phantom{\Big[\Big]} \\
\Upsilon \equiv \delta\varphi + \dot{\varphi}_0 \big(B-\dot{E}\big)
\phantom{\Big[\Big]} &
\Delta \equiv \delta\rho + \dot{\rho}_0 \big(B-\dot{E}\big) 
\,, \phantom{\Big[\Big]}
\end{array}
\label{gaugeinv}
\eeq
where $\varphi_0$ is the homogeneous component of the inflaton field and $\delta\rho$ represents the perturbation in a more generic energy density, $\rho_0$ being the homogeneous component.  We henceforth use the gauge freedom associated with $\bfx$ translation to set $E=0$.  The Einstein field equations then give the constraint 
\beq
\Phi=\Psi \,,
\label{constraint1}
\eeq 
and, on scales larger than the apparent horizon, 
\bea
3\cH^2\Psi + 3\cH\dot{\Psi} &=& -4\pi G\Big[\dot{\varphi}_0\Upsilon
+a^2(dV/d\varphi)\Upsilon-\dot{\varphi}_0^2\Psi\Big] \\
3\cH^2\Psi + 3\cH\dot{\Psi} &=& -4\pi G\,a^2\Delta
\phantom{\Big[\Big]} \label{constraint2} \,.
\eea 

As we have remarked, among the (scalar) perturbations there is one degree of freedom.  A careful exposition of the self-interactions of this degree of freedom is given in \cite{Maldacena:2002vr}, which works in a gauge defined by $\delta\varphi=E=0$.  The scalar degree of freedom can be written in terms of gauge-invariant variables according to 
\beq
\zeta = -\Psi + \frac{\cH^2}{\cH^2-\dot{\cH}}\big(\Phi-\dot{\Psi}) \,.
\label{zeta}
\eeq
The leading-order action for $\zeta$ is that of a free field in de Sitter space, in particular
\beq
S = \frac{1}{2}\int d\eta\,d\bfx\,
\frac{a\,\dot{\varphi}^2_0}{\cH^2}\Big[ \dot{\zeta}^2 - 
(\bm{\partial}\zeta)\!\cdot\!\bm{\partial}\zeta\Big] \,.
\eeq
We use $\zeta$ to establish the quantum theory of the effective Fock space of Section \ref{sec:measures}.  To begin, we perform a mode expansion of $\zeta$.  We refer to both Cartesian mode functions, 
\bea
\zeta_\bfk &\equiv& \int d\bfx\, e^{-i\bfk\cdot\bfx}\,\zeta(\eta_0,\bfx) 
= v_k(\eta_0)\,a_\bfk+v^*_k(\eta_0)\,a^*_{-\bfk} \,,
\label{modes-C}
\eea
and spherically symmetric mode functions
\bea
\zeta_{k\ell m} &\equiv& \int d\Omega_2 \int r^2dr\,
\sqrt{2/\pi}\,k\,j_\ell(kr)\, Y^m_\ell(\Omega_2)\,
\zeta(\eta_0,\bfx) \\
&=& v_k(\eta_0)\,a_{k\ell m} + v^*_k(\eta_0)\,a^*_{k\ell m} \,. 
\phantom{T^{T^{T^{T^T}}}} \label{modes-ss}
\eea
In each case the modes are evaluated at some reference time $\eta=\eta_0$ near the end of inflation.  The $v_k(\eta)$ are taken to satisfy the equation of motion (with $\bm{\partial}\!\cdot\!\bm{\partial}\to -k^2$), given Bunch-Davies boundary conditions, with their normalization set so that the Fourier coefficients $a_\bfk$ and $a^*_{-\bfk}$ ($a_{k\ell m}$ and $a^*_{k\ell m}$) satisfy the usual ladder-operator commutation relations when promoted to quantum operators.  We use the real basis of the spherical harmonics $Y^m_\ell$ ($\Omega_2$ represents the angular coordinates on the 2-sphere), and the $j_\ell$ are spherical Bessel functions.  

The classical perturbation $\zeta$ is promoted to a quantum operator $\hat{\zeta}$ by replacing the classical Fourier coefficients $a^{\phantom{*}}_\bfk$ and $a^*_{-\bfk}$ ($a^{\phantom{*}}_{k\ell m}$ and $a^*_{k\ell m}$) with the ladder operators $\hat{a}^{\phantom{*}}_\bfk$ and $\hat{a}^\dagger_{-\bfk}$ ($\hat{a}^{\phantom{*}}_{k\ell m}$ and $\hat{a}^\dagger_{k\ell m}$).  We promote the corresponding mode functions $\zeta_\bfk$ and $\zeta_{k\ell m}$ similarly.  As a matter of shorthand, we also write other perturbative quantities as quantum operators, for instance the gauge-invariant perturbation $\hat{\Psi}$ or its Fourier mode $\hat{\Psi}_{\!\bfk}$.  It should be understood that these quantities are defined in relation to $\hat{\zeta}$ and $\hat{\zeta}_\bfk$, via multiplicative factors deduced from the gauge transformations and constraint equations given above. 

Although we compute sub-leading effects coming from the fat geodesic and causal patch measures, our calculus does not require using the sub-leading interaction terms in the action to compute them.  Therefore we only need refer to leading-order description outlined above.  Of course, the effects of sub-leading interaction terms compete with the corrections we compute; however these can be found elsewhere in the literature (see for example \cite{Maldacena:2002vr}).      

We now briefly describe the evolution of these perturbations.  It is convenient to do this in longitudinal gauge---defined by setting $B=E=0$---both because in this gauge the metric perturbations are equal to gauge-invariant quantities, $\phi=\Phi=\psi=\Psi$, and because it allows us to draw upon textbook results (see for example \cite{Dodelson:2003ft}).  On scales larger than the apparent horizon, $\Psi$ is roughly constant with respect to time during radiation and matter domination.  Indeed, during matter domination $\Psi$ is approximately constant with respect to time even on scales within the apparent horizon.  During radiation domination, $\Psi$ decays on scales  within the apparent horizon.  The net effect is usually modeled by working in Fourier space and by introducing a transfer function $T(k)$, whereby
\beq
\Psi_{\!\bfk}(\eta_{\rm obs}) \approx \Psi_{\!\bfk}(\eta_{\rm eq}) 
= T(k)\,\Psi_{\!\bfk}(\eta_0) \,,
\label{beth}
\eeq 
with
\beq
T(k) \approx \left\{ 
\begin{array}{ll}
\displaystyle\,\, 1 \phantom{T_{T_{T_{T}}}} \qquad 
& {\rm for}\,\, k\lesssim 1/\rCAH \\
\displaystyle\,\, \frac{\ln(k\rCAH)}{k^2\rCAH^2} \qquad 
& {\rm for}\,\, k\gg 1/\rCAH\,\,, \\
\end{array}
\right. 
\label{transfer}
\eeq
where $\eta_{\rm eq}$ denotes the time of matter / radiation equality and we ignore factors of order unity.  That is, until cosmological-constant domination $\Psi$ is roughly unchanged from its primordial value all the way down to the comoving scale $\rCAH$, below which it decreases with decreasing scale.  During cosmological-constant domination, $\Psi$ decays with time on all scales obeying a linearized analysis.  Solving (\ref{constraint2}) using results from the toy model of Section \ref{ssec:model}, we find  
\beq
\Psi_{\!\bfk}(\eta) = \frac{\eta}{\eta_{\rm obs}}\, 
\Psi_{\!\bfk}(\eta_{\rm obs}) \qquad\,\, 
{\rm for}\,\, \eta\geq\eta_{\rm obs} \,.
\label{PsiafterCC}
\eeq

Our cosmological model takes the density perturbation to be adiabatic.  Thus, on scales larger than the apparent horizon, the total density contrast $\delta\rho/\rho$ is equal to the radiation density contrast $\delta\rho_{\rm rad}/\rho_{\rm rad}$ which is equal to the matter density contrast $\delta\equiv\delta\rho_{\rm m}/\rho_{\rm m}$.  These quantities are constrained by (\ref{constraint2}).  Accordingly, on scales larger than the apparent horizon and at times below the onset of cosmological constant domination, we have  
\beq
\dL = -2\Psi \qquad\,\, \mbox{(beyond the apparent horizon)}\,,
\label{dconstraint}
\eeq
and likewise for $\delta\rho_{\rm rad}/\rho_{\rm rad}$.  We have added the superscript to $\dL$ to avoid confusion with the density contrast computed in another gauge, given below.  It can be shown, by studying the divergencelessness of the stress-energy tensor, that these results continue to hold after cosmological-constant domination, at least on scales larger than the apparent horizon.  

On scales within the apparent horizon, $\dL$ grows with time.  During radiation domination this growth is logarithmic with respect to the growth of the scale factor, while during matter domination the growth is linear with respect to the growth of the scale factor.  The net effect can be modeled in Fourier space, giving
\beq
\dL_\bfk(\eta_{\rm obs}) = 
-\left(2+\gamma k^2\eta_{\rm obs}^2\right)
T(k)\,\Psi_{\!\bfk}(\eta_0) \,.
\label{smallscalematter}
\eeq
On scales larger than the apparent horizon---scales corresponding to $k\gtrsim\rCP=|\eta_{\rm obs}|$ at the time $\eta_{\rm obs}$---the matter contrast is given by its primordial value, reflected by the first term in (\ref{smallscalematter}).  On scales within the apparent horizon, $\dL$ is enhanced by the growth of modes after they enter the horizon, the net effect being modeled by the factor $\gamma k^2\eta^2_{\rm obs}$.  We have included an order-unity factor $\gamma$ to balance the different levels of precision entering the calculations of these two terms.  The sub-horizon evolution of $\dL$ after cosmological-constant domination does not enter into our calculations.

Finally, we return to the radiation contrast $\delta\rho_{\rm rad}/\rho_{\rm rad}$.  Its sub-horizon evolution is more complicated than that of $\Psi$ and $\dL$, however for our analysis is only important to note that it does not grow---aside from an order-unity enhancement on scales corresponding to the first few acoustic peaks in the CMB---and indeed on sufficiently small scales (scales well below $\rCAH$ at the time of last scattering), it decays, due to Silk damping.

\subsection{Conditioning on the average CMB temperature}
\label{sssec:conditions}

In Section \ref{sec:measures}, we conditioned predictions using a given average CMB temperature $T_{\rm obs}$.  This condition is automatically satisfied on a constant-time hypersurface in a gauge that sets the radiation perturbation to zero, $\delta\rho_{\rm rad}=0$.  To keep the analysis simple, we set the gauge so as to satisfy this condition on scales larger than the apparent horizon, and then take this as an approximation for satisfying the condition on all scales.  The accuracy of this approximation is enhanced by the fact that the average CMB temperature is dominated by scales larger than the apparent horizon at last scattering---this because last scattering occurs soon after matter / radiation equality, and $\rCMB\gg\rCAH$---while the sum variance on smaller scales is suppressed, due to statistical regression to the mean.  

To find the appropriate gauge, first note that on scales larger than the apparent horizon, the radiation perturbation is set to zero when the total density perturbation $\delta\rho$ is set to zero.  Since the quantity $\Delta$ in (\ref{gaugeinv}) is gauge-invariant, $\delta\rho$ is set to zero by shifting $B$ by an amount $\delta\rho/\dot{\rho}_0$.  The shift in $B$ is accomplished by an appropriate redefinition of the time parameter, according to (\ref{redefinitions}) and (\ref{newgauges}).  For example, starting in longitudinal gauge, we use
\beq
\sigma = \frac{\delta\rho}{\dot{\rho}_0} = 
-\frac{\dL}{3\cH} = \frac{2}{3}\frac{\Psi}{\cH}\,,
\eeq 
where the the last two expressions incorporate our interest in setting the condition on scales larger than the apparent horizon, during and after matter domination.  This gives
\beq
\phi = \bigg(\frac{1}{3}+\frac{2}{3}\frac{\dot{\cH}}{\cH^2}\bigg)\Psi
-\frac{2}{3}\frac{\dot{\Psi}}{\cH} \,,\qquad
B = \frac{2}{3}\frac{\Psi}{\cH} \,,\qquad
\psi = \frac{5}{3}\,\Psi\,,\qquad
E = 0 \,. \,\,
\label{gaugechoice}
\eeq
While the gauge choice (\ref{gaugechoice}) sets the matter perturbation to zero on scales larger than the apparent horizon, within the apparent horizon the matter perturbation grows relative to its primordial value and is therefore non-zero.  Given the matter contrast in longitudinal gauge, $\dL$, we solve for the density contrast in the $\delta\rho_{\rm rad}=0$ gauge by using the gauge invariance of $\Delta$.  This gives   
\beq
\delta_\bfk(\eta_{\rm obs}) = -\gamma k^2\eta_{\rm obs}^2\,
T(k)\,\Psi_{\!\bfk}(\eta_0) \,.
\label{matterb}
\eeq
Note that while our calculation of $\delta_\bfk(\eta_{\rm obs})$ is simplified by the approximate manner in which we impose the gauge condition $\delta\rho_{\rm rad}=0$, it is not non-zero simply as a consequence of this approximation.  In particular, the radiation perturbation does not significantly increase on scales within the apparent horizon, while the matter perturbation does.  Clearly, (\ref{matterb}) arises entirely as a consequence of the relative growth of the matter contrast.

We emphasize that we use the gauge choice (\ref{gaugechoice}) because in this gauge, hypersurfaces of constant time correspond to surfaces of constant-average-CMB temperature (at our level of approximation).  Insofar as our results depend on gauge-dependent quantities---such as the matter contrast or the three-volume on a fixed-time slice in the causal patch---it is because our calculations depend on the conditional data---the average CMB temperature $T_{\rm obs}$---and not because of residual gauge dependence.  That is, we could perform the calculations in any gauge and get the same results, it is just much simpler to work in the gauge (\ref{gaugechoice}), which makes our conditioning data manifest on constant-time slices.

It is not unexpected that predictions depend on the conditioning data, but this raises the question of what is the correct data to condition on.  To make the most accurate predictions, one should condition on all data of which one is aware.  Yet, as a matter of expediency, or to explain data that one already possesses, one may wish to condition on less \cite{Garriga:2007wz}.  This is the context of our work.  When conditioning on less, it is important to remember that predictions of probabilistic outcomes imply a notion of typicality---that is, it is only meaningful to compare a measured distribution to the predicted distribution if the hypothesis asserts that the measurements are drawn randomly from the predicted distribution---and so the conditional data should be sufficient to select an ensemble from which our measurements can be considered randomly drawn.  While it only increases the accuracy of our prediction to condition on a given average CMB temperature $T_{\rm obs}$, we do not condition on our residing in a galaxy much like the Milky Way (we simply take the number of observers to be proportional to the matter density), which is another important datum.  One could in principle condition on both, but this would be technically challenging and is beyond the scope of this paper.  

Finally, it is worth noting that while these considerations are important for the physical interpretation of our calculations, the actual numerical order of magnitude of our results is insensitive to the precise details of the conditioning data that is employed.


\section{Correlation functions in the fat geodesic measure}
\label{sec:FG}

Section \ref{sec:measures} demonstrates that the fat geodesic and the causal patch measures in general predict different values for inflationary observables.  Furthermore, these predictions differ from those of the standard formulation of inflationary predictions.  The differences are apparently small.  To better appreciate the size and form of the measure selection effects, we here compute expectation values for correlation functions of the Fourier components of the gauge-invariant Newtonian potential $\Psi$.  While these are not actually physical observables---in the context of CMB measurements, the physical observables relate to temperature variations with respect to position in the sky of the temperature of photons, which arrive at a local detector after free-streaming from the surface of last scattering---there is nothing new in converting the correlation functions of $\Psi$ into a spectrum of temperature fluctuations on a local sky (see for example \cite{Hu:1997mn}).  Therefore, we set these complications aside.    

The analysis of Section \ref{sec:measures} is straightforward to apply to the fat geodesic measure.  To begin, consider the expectation value of the 2-point correlation function of the Fourier modes $\Psi_{\!\bfk}$.  According to the result (\ref{FGresult}), this can be written
\bea
\<\Psi_{\!\bfk}\,\Psi_{\!\bfk'}\>_{\rm obs} &=& 
\frac{\<0|\overline{\rho}_{\rm m}^2(\eta_{\rm obs})
\big[1+\hat{\delta}(\eta_{\rm obs},0)\big]^2
\hat{\Psi}_{\!\bfk}\,\hat{\Psi}_{\!\bfk'}|0\>}
{\<0|\overline{\rho}_{\rm m}^2(\eta_{\rm obs})
\big[1+\hat{\delta}(\eta_{\rm obs},0)\big]^2|0\>} \\
&=& \frac{\<0|\big[1+2\hat{\delta}(\eta_{\rm obs},0)\big]
\hat{\Psi}_{\!\bfk}\,\hat{\Psi}_{\!\bfk'}|0\>}
{\<0|\big[1+2\hat{\delta}(\eta_{\rm obs},0)\big]|0\>} \,,
\label{FG2-point1}
\eea
where $\overline{\rho}_{\rm m}$ is the homogeneous component of the matter density, which cancels between the numerator and denominator, and we have expanded perturbatively in the matter contrast $\delta$.  The expression still looks complicated but, given the ladder-operator algebra, these quantities are easy to compute.  Only even factors of field insertions survive the expectation value in the Bunch-Davies vacuum, so the normalization factor can always be ignored (after accounting for the factor of $\overline{\rho}^2_{\rm m}$).  In the case of the 2-point function, the term involving $\hat{\delta}$ can also be ignored, and we recover the standard prediction:
\beq
\<\Psi_{\!\bfk}\,\Psi_{\!\bfk'}\>_{\rm obs} = P(k)\,(2\pi)^3\delta(\bfk+\bfk')\,,
\label{power}
\eeq
where $P(k)\propto T^2(k)\,|v_k|^2$ is the power spectrum of $\Psi$.  Note that, to keep expressions simple, we include in $P(k)$ the effects of the transfer function on scales $k\ll1/\rCAH$.  Aside from that, $P(k)$ is proportional to $1/k^3$ times a weak function of $k$, which depends on the details of the inflationary cosmology (but not on the measure, given our conditioning assumptions).  

Although the fat geodesic measure does not modify the inflationary prediction for the 2-point correlator, it is now clear that we should look at the 1-point and 3-point correlation functions.  Indeed, we find the 1-point expectation value
\bea
\<\Psi_{\!\bfk}\>_{\rm obs} &=& 
\<0|\big[1+2\hat{\delta}(\eta_{\rm obs},0)\big]
\hat{\Psi}_{\!\bfk}|0\> \phantom{\Big[\Big]} \\
&=& -2\,\<0|\int\frac{d\bfq}{(2\pi)^3}\,\gamma q^2\eta_{\rm obs}^2
\hat{\Psi}_{\!\bfq}\,\hat{\Psi}_{\!\bfk}|0\> \\
&=& -2\gamma k^2\eta_{\rm obs}^2\, P(k) \,, 
\phantom{\Big[\Big]}
\label{mean1}
\eea
where we have used (\ref{matterb}) to express the matter contrast $\delta$ in terms of $\Psi$, evaluating $\delta$ in the gauge (\ref{gaugechoice}) to automatically incorporate the fixed-average-CMB-temperature condition.  The 1-point expectation value is usually zero by construction.  However, in the present case each perturbation $\Psi$ is defined against its `mean' value (on the $\eta=\eta_0$ hypersurface in a given bubble) before the fat geodesic of the bubble is weighted according to the measure.  When we subsequently weight the fat geodesic of the perturbation $\Psi$ according to the measure, we find the expectation value of $\Psi$ is shifted relative to this `prior' mean.  If we were computing statistics of $\Psi$ in position space, we might first subtract off the mean value of (the observed portion of) the perturbation $\Psi$.  Since we work in Fourier space, the spatially-averaged mean affects only the $\bfk=0$ mode, which has already been discarded.  We are left with the scale-dependence of the shift, given by (\ref{mean1}).

The 1-point expectation value (\ref{mean1}) takes the form of a monopole.  This could have been guessed based on the (statistical) symmetries of the physical scenario, and is evinced by the fact that (\ref{mean1}) depends only on the magnitude of $\bfk$.  It is manifest when we express the 1-point function in terms of the spherically symmetric mode functions,
\bea
\<\Psi_{\!k\ell m}\>_{\rm obs} &=& 
-2\,\<0|\int dq\sum_{\ell',m'}\sqrt{2/\pi}\,q\,
j_{\ell'}(0)\,Y^{m'}_{\ell'}\!(0)\,\gamma q^2\eta_{\rm obs}^2
\hat{\Psi}_{\!q00}\,\hat{\Psi}_{\!k\ell m}|0\> \\
&=& -\big(\sqrt{2}/\pi\big)\,
\gamma k^3\eta_{\rm obs}^2 P(k)\,
\delta_{\ell0}\,\delta_{m0} \,, 
\phantom{\Big[\Big]}
\label{mean1b}
\eea
where we have input $j_\ell(0)=\delta_{\ell0}$ and $Y^0_0=(4\pi)^{-1/2}$.  The extra power of $k$ relative to (\ref{mean1}) expresses the different measure on radial modes between Cartesian and spherically symmetric coordinates.  While the scale-dependence of a monopole mode is in principle detectable in the CMB---it affects the apparent scale of the acoustic peaks, for example---its imprints are subdominant.  As a qualitative guide, one can liken the observability of the monopole to the observability of spatial curvature in an open (FRW) cosmology. 

To appreciate the size of the measure selection effect, first note that the scale of observable wavenumbers is set by the comoving size of the surface of last scattering, $\rCMB$, in particular observable scales correspond to $k\gtrsim 1/\rCMB$.  Meanwhile, $|\eta_{\rm obs}|=(1/2)\,\rCMB$.  Therefore, the factors of $k$ provides some enhancement of the monopole on scales of interest.  On the other hand, the size of the 1-point function is always below cosmic variance.  This can be seen by simply noting the size and scale dependence of a typical $\Psi$ perturbation, 
\beq
\Psi_{\!\bfk}\sim \big[P(k)\big]^{1/2} 
\sim k^{-3/2}\, \big[k^3P(k)\big]^{1/2} \,.
\eeq 
Up to slow-roll corrections and a loss of power on comoving scales below $\rCAH$, the factor in brackets in the second expression is dimensionless, independent of $k$, and of order $10^{-5}$.  Our result (\ref{mean1}) differs by a factor of order $k^{1/2}\big[k^3P(k)\big]^{1/2}$.  The term in brackets provides an order $10^{-5}$ suppression.  To compensate for this suppression with the factor of $k^{1/2}$ would require examining extremely small scales, $k\rCMB\sim 10^{10}$, well beyond the validity of our analysis and beyond any foreseeable observation.

Finally, we turn to the 3-point correlation function.  It is given by
\beq
\<\Psi_{\!\bfk}\,\Psi_{\!\bfk'}\,\Psi_{\!\bfk''}\>_{\rm obs} =
-2\gamma k^2\eta_{\rm obs}^2
P(k)P(k')\,(2\pi)^3\delta(\bfk'+\bfk'') + {\rm perms}\,,
\eeq
or, in terms of spherical harmonics,
\bea
\<\Psi_{\!k\ell m}\,\Psi_{\!k'\ell'm'}\,\Psi_{\!k''\ell''m''}\>_{\rm obs} 
&=& -\big(\sqrt{2}/\pi\big)\,\gamma k^3\eta_{\rm obs}^2 P(k)P(k')\, 
\delta(k'-k'')\,\delta_{\ell'\ell''}\,\delta_{m'm''}\,
\delta_{\ell0}\,\delta_{m0} \qquad\, \nn\\
& & +\, {\rm perms} \,. 
\phantom{T^{T^{T^T}}} 
\label{3point1}
\eea
Notice that in each term of the sum, one of the wavenumbers is unconstrained.  Usually, the 3-point function is proportional to a total-wavenumber-conserving delta function; for example in Cartesian coordinates there is a factor of $\delta(\bfk+\bfk'+\bfk'')$.  This factor arises because the usual 3-point correlator involves field insertions coming from an expansion in the local self-interaction terms of the sub-leading action, and the local interactions conserve momentum.  The 3-point correlator (\ref{3point1}) comes from a non-local correlation between the matter contrast $\delta$ and the perturbation mode $\Psi_{\!\bfk}$---a consequence of the measure---therefore there is no reason to expect total wavenumber conservation.  

The 3-point function is associated with non-Gaussianity.  This is usually parametrized in terms of a quantity $f_{\rm NL}$ \cite{Bartolo:2004if}, which in terms of $\Psi$ would roughly correspond to 
\beq
\<\Psi_{\!\bfk}\,\Psi_{\!\bfk'}\,\Psi_{\!\bfk''}\>\sim 
f_{\rm NL}\,\big[ P(k)P(k') + {\rm perms.} \big] \,
\delta(\bfk+\bfk'+\bfk'')\,.
\eeq
Since the source of non-Gaussianity from the measure selection effect is non-local, we cannot draw a direct connection to the parameter $f_{\rm NL}$.  Nevertheless, setting aside the delta functions, it appears as if the fat-geodesic measure predicts non-Gaussianity of a magnitude in rough correspondence with $f_{\rm NL}\sim \gamma k^2\rCMB^2$.  This is actually larger than the standard slow-roll result \cite{Maldacena:2002vr}, which does not feature the rising $k$ dependence and is also suppressed by a slow-roll parameter.  However, expanding our attention to include the delta functions, we see that the present non-Gaussianity arises from correlations involving the monopole; specifically one of the modes must have $\ell=m=0$.  This alone would seem to make it indiscernible in the CMB, where the effects of the monopole are indirect.  Yet, even in a map of perturbations with more three-dimensional resolution, we expect the 3-point correlator to fall within cosmic variance, since restricting to the single $\ell=m=0$ mode does not allow one to reduce the uncertainty as is usually done by averaging over many $m$ at large $\ell$.


\section{Correlation functions in the causal patch measure}
\label{sec:CP}

\subsection{Observables and operators}

The analysis of Section \ref{sec:measures} describes how to make predictions using the causal patch measure, however the main result (\ref{CPprediction}) is not expressed in a form that is amenable to predicting the expectation values of correlation functions.  We here construct a more tractable expression for the operator $\hat{Z}$ that appears in (\ref{Zdef}).

The operator $\hat{Z}$ depends on the physical observable $z$ that one wishes to predict; we are interested in expectation values of correlation functions involving products of the Cartesian Fourier modes $\Psi_{\!\bfk}$.  Note that these quantities are independent of the location in the causal patch at which they are measured.\footnote{The Fourier modes $\Psi_{\!\bfk}$ are not directly observable, while the actual observables---the spherical harmonic coefficients $a_{\ell m}$ of temperature variations in CMB photons---depend on the location at which they are observed.  However, the expectation values of products of $a_{\ell m}$ can be written as functionals of the expectation values of products of $\Psi_{\!\bfk}$, where the latter are computed as if the $\Psi_{\!\bfk}$ were observable.  The same applies here, except the expectation values of products of $\Psi_{\!\bfk}$ are computed with the measure, as opposed to as straight correlation functions.  In the end these are the quantities that we compute, and they are independent of the location at which they are measured.}  This means that they factor out of the volume integral of $\hat{Z}$, which makes it worthwhile to study the operator $\hat{Z}$ but with the various insertions of $\Psi_{\!\bfk}$ factored out---an operator that we denote $\hat{Z}_0$---inserting the factors of $\Psi_{\!\bfk}$ back into $\hat{Z}$ nearer to the end of the calculation.  

The volume integral of $\hat{Z}_0$ integrates over both space and time, but with a delta function selecting for the time at which the average CMB temperature attains the given value $T_{\rm obs}$.  This delta function is automatically satisfied (at our level of approximation) on an appropriate fixed-time hypersurface in the gauge (\ref{gaugechoice}).  Therefore, in this gauge we can write 
\beq
\hat{Z}_0 \propto \int_{\rm CP[\Psi]} \sqrt{-g[\eta_{\rm obs},\Psi]}\,
d\bfx\,\,\overline{\rho}_{\rm m}^2(\eta_{\rm obs})
\big[1+\hat{\delta}(\eta_{\rm obs},0)\big]
\big[1+\hat{\delta}(\eta_{\rm obs},\bfx)\big] \,.
\eeq
The rotational symmetry of the causal patch makes the spherically symmetric mode functions most convenient.  Expanding to linear order in perturbative quantities, we write
\bea
\hat{Z}_0 &\propto& \int d\Omega_2 \int_0^{\tr(\Omega_2)}
r^2dr\,\bigg[ 1 + \int dq\sum_{\ell,m}\sqrt{2/\pi}\,q\,
j_\ell(qr)\,Y^m_\ell(\Omega_2)\!
\left( \hat{\delta}_{q\ell m} - 5\hat{\Psi}_{\!q\ell m}\right) 
\bigg. \nn\\
& & +\, \bigg. \int dq\sum_{\ell,m}\sqrt{2/\pi}\,q\,
j_\ell(0)\,Y^m_\ell(0)\,\hat{\delta}_{q\ell m} \bigg] \,,
\eea
where $\tr$ is the comoving radius of the causal patch, which in the presence of metric perturbations depends on the angular coordinates.  We have dropped the factor of $a^3(\eta_{\rm obs})\,\overline{\rho}_{\rm m}^2(\eta_{\rm obs})$ and have expressed the metric perturbation $\psi$ in terms of $\Psi$ using (\ref{gaugechoice}), delaying the corresponding substitution with $\delta$ to keep the expressions simpler.  The radial integration can be performed for all terms, but the expressions are rather complicated for some of them.  In the end, the details are not important, so we simply define
\beq
I_{q\ell}(\tr) \equiv \int_0^{\tr} r^2dr\,q^3\,j_\ell(qr) \,.
\eeq
Performing the radial integration, we obtain
\bea
\hat{Z}_0 &\propto& \int d\Omega_2\, \bigg[ \frac{1}{3}\,
\tr^3(\Omega_2) +\int \frac{dq}{q^2}\sum_{\ell,m}\sqrt{2/\pi}\,
I_{q\ell}\big[\tr(\Omega_2)\big]\,
Y^m_\ell(\Omega_2)\!
\left( \hat{\delta}_{q\ell m} - 5\hat{\Psi}_{\!q\ell m}\right) 
\bigg. \nn\\*
& & +\,\bigg. \frac{1}{3}\,\tr^3(\Omega_2) 
\int \frac{q\,dq}{\sqrt{2}\pi}\,\hat{\delta}_{q00} \bigg] \,.
\label{Z00}
\eea

Note that while our analysis does not describe the perturbations $\delta$ and $\Psi$ down to arbitrarily small scales, we have nevertheless integrated over all scales in the causal patch.  Since the integrations were performed in Fourier space, the analysis is accurate so long as we do not re-sum the Fourier mode expansion and so long as we restrict attention to scales on which the perturbations $\delta$ and $\Psi$ are accurately described by the mode expansion.  Since in the end the observables we are interested in are statistics of Fourier components of $\Psi$, we only re-sum the modes on the scales corresponding to those Fourier components.  Meanwhile, although the description of the evolution of modes within the apparent horizon in Section \ref{ssec:gauge} is not precise, it is accurate at an order-of-magnitude level over scales well below $\rCAH$, which is very small next to $\rCMB$, which is of order $\rCP$.  Therefore, there are plenty of modes on scales for which our integration over the causal patch is accurate.

The comoving distance to the boundary of the causal patch $\tr$ is obtained by solving for the trajectory of a radial null ray backwards from future infinity.  The null condition is 
\beq
-(1+2\phi)\,d\eta^2 + 2\partial_rB\,d\eta\,dr + (1-2\psi)\, dr^2 =0\,.
\label{null1}
\eeq
Since we integrate from future infinity to $\eta_{\rm obs}$, we are interested in the behavior of the metric perturbations during cosmological-constant domination.  In the gauge (\ref{gaugechoice}), this gives
\beq
\phi = \frac{5}{3}\frac{\eta}{\eta_{\rm obs}}\,\Psi\,,\qquad
B = -\frac{2}{3}\frac{\eta^2}{\eta_{\rm obs}}\,\Psi\,, \qquad
\psi = \frac{5}{3}\frac{\eta}{\eta_{\rm obs}}\,\Psi\,, \quad
\eeq
where we have used (\ref{PsiafterCC}), along with $\cH=-1/\eta$ during cosmological-constant domination in our toy cosmological model, and where $\Psi$ is understood to be evaluated at $\eta_{\rm obs}$.  Using the quadratic formula and expanding in perturbations, the null condition (\ref{null1}) becomes  
\beq
-d\eta = \left( 1 + \frac{2}{3}\frac{\eta^2}{\eta_{\rm obs}}\,
\partial_r\Psi - \frac{10}{3}\frac{\eta}{\eta_{\rm obs}}\,
\Psi \right) dr \,.
\label{null2}
\eeq
Simple as it looks, this equation does not permit a separation of variables.  Nevertheless, it is apparent that the solution takes the form $r = -\eta + {\cal O}(\Psi)$.  We can use this to convert the terms in parentheses into functions of $\eta$, then expand in $\Psi$, then perform a mode expansion of $\Psi|_{r\to-\eta}$, then integrate over $\eta$.  This gives (after integration by parts)
\beq
\tr(\Omega_2) = \rCP + \int dq\sum_{\ell,m}\sqrt{2/\pi}\,
\Psi_{\!q\ell m}\,\bigg[\frac{2}{3}q\rCP\,j_\ell(q\rCP)
+\frac{2J_{q\ell}(\rCP)}{q\rCP}\bigg]\, Y^m_\ell(\Omega_2) \,,
\label{rcp}
\eeq
where $\rCP = -\eta_{\rm obs}$ is the previously-defined average comoving distance to the boundary of the causal patch (at the time $\eta=\eta_{\rm obs}$), and we have defined 
\beq
J_{q\ell}(\rCP) \equiv \int_0^{\rCP} rdr\,q^2\,j_\ell(qr) \,.
\eeq

We proceed by inserting the expression for $\tr$ (\ref{rcp}) into expression for $\hat{Z}_0$ (\ref{Z00}).  Expanding in the perturbations, the angular integration can be performed, and we obtain
\bea
\hat{Z}_0 &\propto&  \frac{4\pi}{3}\rCP^3+\sqrt{8}\,\rCP^3\int q\,dq\,\bigg\{
\bigg[\frac{2}{3}\,j_0(q\rCP)+2\,\frac{J_{q0}(\rCP)}{q^2\rCP^2} 
-5\,\frac{I_{q0}(\rCP)}{q^3\rCP^3}\bigg]\hat{\Psi}_{\!q00} \bigg. \nn\\*
& & +\,\bigg. \bigg[1+\frac{I_{q0}(\rCP)}{q^3\rCP^3}\bigg]\hat{\delta}_{q00}\,\bigg\} \,.
\label{Z01}
\eea
Note that the angular integrations kill all of the modes except for the monopole, $\ell=m=0$.

So far it has been convenient to work in terms of spherically symmetric mode functions, but we are interested in predicting expectation values for correlation functions of Cartesian Fourier modes.  (It is important to use Cartesian Fourier modes because, unlike the spherically symmetric mode functions, they do not depend on the origin of coordinates.)  To express $\hat{Z}_0$ in terms of Fourier mode functions, we first note that
\bea
\Psi_{\!q00} &=& \int d\Omega_2 \int r^2dr\,\sqrt{2/\pi}\,q\,
j_0(qr)\,Y^0_0(\Omega_2)\,\Psi(\eta_{\rm eq},\bfx) \nn\\
&=& \int d\Omega_2 \int \frac{rdr}{\sqrt{2}\pi}\,
\sin(qr) \int \frac{d\bfp}{(2\pi)^3}\,e^{i\bfp\cdot\bfx}\,
\Psi_{\!\bfp} \,,
\eea
where in the second line we have substituted for the particular spherically symmetric mode functions and have inserted the Fourier mode expansion for the perturbation $\Psi$.  To simplify this expression, we write $\bfp=p\,\hat{\bfp}$, where $p$ is the magnitude of $\bfp$, and we align the coordinate systems so that $\hat{\bfx}\cdot\hat{\bfp}=\cos(\theta)$, where $d\Omega_2=\sin(\theta)\,d\theta\,d\phi$ (and $\bfx=r\,\hat{\bfx}$).  We can then perform the first angular integral, giving
\beq
\Psi_{\!q00} = \sqrt{2} \int dr\,\sin(qr) 
\int \frac{d\bfp}{(2\pi)^3}\frac{1}{p}\,\sin(pr)\,\Psi_{\!\bfp} \,.
\eeq
We now recognize the integral over $r$ in the context of the orthonormality condition of spherical Bessel functions:
\beq
\int r^2dr\,j_0(qr)\,j_0(pr) = \int dr \sin(qr)\sin(pr) = \frac{\pi}{2q^2}\,\delta(p-q) \,.
\eeq
Plugging into (\ref{Z01}), we obtain our final expression for $\hat{Z}_0$:
\beq
\hat{Z}_0 \propto  \frac{4\pi}{3}\rCP^3+\rCP^3\int\frac{d\bfq}{(2\pi)^2}
\bigg\{\bigg[\frac{2}{3}\,j_0(q\rCP)+2\,\frac{J_{q0}(\rCP)}{q^2\rCP^2} 
-5\,\frac{I_{q0}(\rCP)}{q^3\rCP^3}\bigg]\hat{\Psi}_{\!\bfq} +
\bigg[1+\frac{I_{q0}(\rCP)}{q^3\rCP^3}\bigg]\hat{\delta}_\bfq \bigg\} \,.
\label{Z02}
\eeq

\subsection{Correlation functions}

With the expression (\ref{Z02}) in hand, the calculation of expectation values of correlation functions is straightforward, if sometimes a bit messy.  Following the treatment of the fat geodesic measure in Section \ref{sec:FG}, we first consider the 2-point correlation function.  It is
\beq
\<\Psi_{\!\bfk}\,\Psi_{\!\bfk'}\>_{\rm obs} = 
\frac{\<0|\hat{Z}_0\,\hat{\Psi}_{\!\bfk}\,\hat{\Psi}_{\!\bfk'}|0\>}
{\<0|\hat{Z}_0|0\>} = P(k)\,(2\pi)^3\delta(\bfk+\bfk') \,.
\label{CP2-point1}
\eeq   
The first expression restates (\ref{CPprediction}); the second expression follows from inserting (\ref{Z01}), noting that only even-numbered products of ladder operators give non-zero expectation values in the Bunch-Davies vacuum (and thus the non-trivial terms in $\hat{Z}_0$ do not contribute).  We see that, as in the case of the fat geodesic measure, the causal patch measure does not modify the standard inflationary prediction at leading or first sub-leading order.  

Meanwhile, for the 1-point function, we find 
\beq
\<\Psi_{\!\bfk}\>_{\rm obs} = -\frac{3}{2}\,\bigg\{\!
\gamma k^2\rCP^2 \bigg[1+\frac{I_{k0}(\rCP)}{k^3\rCP^3}\bigg] 
-\frac{2}{3}\,j_0(k\rCP) - 2\,\frac{J_{k0}(\rCP)}{k^2\rCP^2}
+5\,\frac{I_{k0}(\rCP)}{k^3\rCP^3} \bigg\}\, P(k) \,,
\label{CP1-point}
\eeq
where we have used (\ref{matterb}) with $|\eta_{\rm obs}|=\rCP$.  As with the fat geodesic measure, the 1-point expectation value takes the form of a monopole.  (Again, this could have been guessed based on the symmetries of the physical scenario.)  The functions $I_{k0}(\rCP)$, $J_{k0}(\rCP)$, and $j_0(k\rCP)$ are at most of order unity for all values of $k\rCP$.  Note also that $\rCP = (1/2)\,\rCMB$, with observable scales corresponding to $k\gtrsim 1/\rCMB$.  Therefore, the size of the 1-point expectation value is of the same order as the 1-point expectation value for the fat geodesic measure.  In particular, it is well within cosmic variance, despite the enhancement of the first term for large $k$.

While the computations leading the various terms in (\ref{CP1-point}) are somewhat complicated, it is evident that several effects contribute to the final result.  As with the fat geodesic measure, there is an effect coming from the tendency of the worldline defining the measure to gravitate toward over-densities.  Also as with the fat geodesic measure, the calculation incorporates an anthropic effect, whereby the probability of a measurement is taken to be proportional to the matter density; however unlike in the fat geodesic measure in the causal patch measure this factor is integrated over the entire causal patch.  And finally, this integration over the causal patch depends on the total volume in the causal patch, which depends on the metric and its perturbations.  

Finally, we turn to the 3-point correlation function.  It can be written 
\beq
\<\Psi_{\!\bfk}\,\Psi_{\!\bfk'}\,\Psi_{\!\bfk''}\>_{\rm obs} =
\<\Psi_{\!\bfk}\>_{\rm obs}\,P(k')\,
(2\pi)^3\delta(\bfk'+\bfk'') + {\rm perms}\,,
\eeq
where we refer to the result (\ref{CP1-point}) to avoid repeating the long expression.  Each term depends only on the magnitude of one of the wavenumbers $\{\bfk,\bfk',\bfk''\}$ and so, as with the fat geodesic measure, the 3-point function is only non-zero when one of the field insertions is a monopole.  The overall magnitude is also the same as in the expectation value of the 3-point function in the fat geodesic measure, so all of the comments there apply here.


\section{Conclusions}
\label{sec:conclusions}

We have computed the spectrum of inflationary perturbations in the context of the fat geo\-desic and causal patch measures.  Both measures predict a 1-point expectation value for the gauge-invariant Newtonian potential $\Psi$.  This takes the form of a scale-dependent monopole, $\<\Psi_{\!k\ell m}\>_{\rm obs} =\overline{\Psi}_{\!k00}\,k\,\delta_{\ell0}\,\delta_{m0}$, where the precise form of $\overline{\Psi}_{\!k00}$ differs between the two measures, but in both cases the dominant behavior is $\overline{\Psi}_{\!k00}\sim k^2\rCMB^2\,P(k)$, where $\rCMB$ is the comoving size of the surface of last scattering and $P(k)$ is the power spectrum of $\Psi$.  Each measure also predicts a contribution to the expectation value of the 3-point correlation function, when one of the three field insertions is the monopole $\Psi_{\!k00}$, this effect evidently due to correlations with the background $\overline{\Psi}_{\!k00}$.  In each case the prediction is well within cosmic variance.

These predictions take the local cosmological model, including the model of inflation, to be completely specified.  A more general approach would survey the landscape of vacua in the theory and take into account all of the models of inflation (and of the subsequent cosmology) that are consistent with our knowledge of the environment.  Consequently, the more general predictions would include distributions for the amplitude of the perturbations, the tilt, etc., and the profiles of these distributions would depend on the choice of measure.  While this suggests interesting possibilities for testing phenomenology, we stress that our analysis is orthogonal to these considerations.  Indeed, we never actually specified the model of inflation.  The inflaton model parameters determine properties of the power spectrum $P(k)$, however our analysis is independent of the precise form of this function.   
 
We can now rephrase the analogy of the introduction, involving darts being thrown at a wall, in terms of these worldline-based measures.  Consider the worldline in the bubble-nucle\-ation geometry illustrated in Figure \ref{fig:bubble}.  As it enters the bubble, the worldline is not comoving with respect to the open-FRW frame in the bubble.  Although the worldline quickly becomes comoving, the initial misalignment of comoving frames gives what might be seen as a large effect:  surveying all future histories of the worldline---over which the bubble nucleates, with uniform frequency, at all points in the local spacetime---the worldline overwhelmingly passes through points in the bubble within a few curvature radii of the center of the bubble.  (Here, the center of the bubble is defined by the comoving worldline that passes through the center of the bubble-nucleation event.)  This effect might be seen as large because the three-volume on constant-time hypersurfaces in any given bubble diverges.  That is, based on the symmetries, one might naively conclude that there is zero probability to lie within a few curvature radii of the center.  On the other hand, the FRW symmetries in the bubble makes it difficult for an observer to discern his distance from the center.  Nevertheless, given some small perturbation to this background, we recover the effect.  Specifically, the worldline, which becomes comoving with respect to surfaces of uniform energy density, is not comoving with respect to surfaces of uniform radiation density (after structure formation), and meanwhile an observer can discern his location in relation to the perturbation.  This is the source of one measure selection effect on the observed inflationary spectrum.  The size of the effect is, roughly speaking, suppressed by the smallness of the perturbation.  With the causal patch measure, there is also an effect coming from the size of the causal patch.

In the introduction we mentioned a related analysis with respect to the proper-time cutoff measure.  In that case the scale-dependent monopole is expected to be much more significant than with the fat geodesic and causal patch measures.  We believe this is due to the youngness problem of the proper-time cutoff measure, by which the measure gives overwhelming weight to regions featuring perturbations that would be on the extreme tail of the (approximately) Gaussian distribution predicted by assuming global FRW symmetries.  Thus, while the size of the effect is `suppressed' by the size of the perturbations, the measure selects for unusually large perturbations and so the size of the effect is significant.  

One can view selection effects associated with the choice of measure as a consequence of broken FRW symmetries on the largest scales.  In light of this, one might be tempted to view the choice of measure as an infrared effect, providing corrections that can be parametrized in terms of the usual tools of effective field theory.  While this view might have some (limited) validity, we have yet to find a crisp formulation of it.  The immediate problem with the usual approach is we do not know our location in the global spacetime---meaning we must consider that we could be at any point consistent with our knowledge of the local environment---which forces us to consider the global properties of spacetime even as we focus on local observations.  And indeed, while the measure effects studied in this paper are small, they are suppressed in terms of the size of the inflationary perturbations, not in terms of any infrared cosmological scale.  This failure of effective-field-theoretic intuition is more dramatic in light of the three phenomenological tests mentioned in Section \ref{sec:introduction}, which demonstrate how the choice of measure can have very large effects.  Therefore, we consider it an important proof of principle that the fat geodesic and causal patch measures validate the standard inflationary predictions, up to effects within cosmic variance.

\acknowledgments

The author thanks Dionysios Anninos, Raphael Bousso, Daniel Harlow, Steve Shenker, and Douglas Stanford for many invaluable discussions.  The author is also grateful for the support of the Stanford Institute for Theoretical Physics.


\providecommand{\href}[2]{#2}\begingroup\raggedright\endgroup

\end{document}